\documentclass[iop]{emulateapj}
\usepackage{apjfonts}

\usepackage{multirow}
\usepackage{longtable}
\usepackage{natbib}
\usepackage{color}

\slugcomment{Draft \today}

\shorttitle{Optical Counterparts to Four Energetic {\em Fermi} MSPs}
\shortauthors{Breton et al.}

\begin{document}

\title{Discovery of the Optical Counterparts to Four Energetic {\em Fermi} Millisecond Pulsars}

\author{R. P. Breton\altaffilmark{1,2}, M. H. van Kerkwijk\altaffilmark{2}, M. S. E. Roberts\altaffilmark{3,4}, J. W. T. Hessels\altaffilmark{5,6}, F. Camilo\altaffilmark{7,8}, M. A. McLaughlin\altaffilmark{9}, S. M. Ransom\altaffilmark{10}, P. S. Ray\altaffilmark{11}, I. H. Stairs\altaffilmark{12}}

\altaffiltext{1}{School of Physics and Astronomy, University of Southampton, Southampton, SO17 1BJ, UK; r.breton@soton.ac.uk.}
\altaffiltext{2}{Department of Astronomy and Astrophysics, University of Toronto, Toronto, ON M5S 3H4, Canada.}
\altaffiltext{3}{Eureka Scientic Inc., 2452 Delmer Street, Suite 100, Oakland, CA 94602-3017, USA.}
\altaffiltext{4}{Ithaca College, 205 Williams Hall, Ithaca, NY 14850, USA.}
\altaffiltext{5}{ASTRON, the Netherlands Institute for Radio Astronomy, Postbus 2, 7990 AA, Dwingeloo, The Netherlands.}
\altaffiltext{6}{Astronomical Institute ``Anton Pannekoek'', University of Amsterdam, Science Park 904, 1098 XH Amsterdam, The Netherlands.}
\altaffiltext{7}{Columbia Astrophysics Laboratory, Columbia University, 550 West, 120th Street, New York, NY 10027, USA.}
\altaffiltext{8}{Arecibo Observatory, HC3 Box 53995, Arecibo, PR 00612, USA.}
\altaffiltext{9}{Department of Physics, White Hall, West Virginia University, Morgantown, WV 26506, USA.}
\altaffiltext{10}{National Radio Astronomy Observatory, Charlottesville, VA 22903, USA.}
\altaffiltext{11}{Space Science Division, Naval Research Laboratory, Code 7655, 4555 Overlook Avenue SW, Washington, DC 20375, USA.}
\altaffiltext{12}{Department of Physics and Astronomy, University of British Columbia, 6224 Agricultural Road, Vancouver, BC V6T 1Z1, Canada.}

\begin{abstract}

In the last few years, over 43 millisecond radio pulsars have been discovered by targeted searches of unidentified $\gamma$-ray sources found by the {\em Fermi Gamma-Ray Space Telescope}. A large fraction of these millisecond pulsars are in compact binaries with low-mass companions. These systems often show eclipses of the pulsar signal and are commonly known as {\em black widows} and {\em redbacks} because the pulsar is gradually destroying its companion. In this paper, we report on the optical discovery of four strongly irradiated millisecond pulsar companions. All four sources show modulations of their color and luminosity at the known orbital periods from radio timing. Light curve modelling of our exploratory data shows that the equilibrium temperature reached on the companion's dayside with respect to their nightside is consistent with about $10-30$\% of the available spin-down energy from the pulsar being reprocessed to increase the companion's dayside temperature. This value compares well with the range observed in other irradiated pulsar binaries and offers insights about the energetics of the pulsar wind and the production of $\gamma$-ray emission. In addition, this provides a simple way of estimating the brightness of irradiated pulsar companions given the pulsar spin-down luminosity. Our analysis also suggests that two of the four new irradiated pulsar companions are only partially filling their Roche lobe. Some of these sources are relatively bright and represent good targets for spectroscopic follow-up. These measurements could enable, among other things, mass determination of the neutron stars in these systems.

\end{abstract}


\section{Introduction}\label{s:introduction}

The {\em Fermi Gamma-Ray Space Telescope} (hereafter simply {\em Fermi}) has discovered hundreds of unidentified point sources whose positional uncertainties are small enough to enable deep targeted pulsar searches with the world's largest radio telescopes. These searches have resulted in the discovery of 43 energetic millisecond pulsars (MSPs) to date \citep{cogn11a,hess11a,kerr12a,rans11a,ray12b}. Intriguingly, many of these are ``black widow'' and ``redback'' systems, of which previously only a handful were known -- {\em Fermi} has so far increased their number to over twenty \citep{robe12a}.

Black widows are energetic MSPs (spin-down luminosity $\dot{E} \sim 10^{34-35}$\,erg s$^{-1}$) with very low mass (few $0.01\,M_\odot$) companions in compact, few-hour orbits. The original black widow, PSR~B1957+20 \citep{fruc88b}, and other members of the class derive their name from the fact that their strong relativistic wind ablates the surface of the companion star, which might have been significantly more massive in the past \citep{fruc88b}. Redbacks, on the other hand, have similar pulsars but more massive companions, of $\sim\!0.2\,M_\odot$. As yet, is it not clear whether redbacks are the close progeny of accreting X-ray systems, nor whether they can evolve into black widow systems. The evidence that the prototype redback PSR~J1023+0038 accreted matter not long ago is suggestive of the former at least \citep{arch09a}.

Because of their ``cannibalistic'' behavior, it has been proposed that some black widows eventually completely ablate their companion \citep{kluz88a,van-88a}. If this is the case, it could at least partially resolve the conundrum of the existence of isolated MSPs, since these sources clearly appear to have evolved in binary systems despite the fact that they are now isolated. It appears, however, that at least for the original black widow pulsar, PSR~B1957+20, the evaporation is too slow \citep{eich88a,eich95a}. Hence one has to wonder whether the known black widows are somewhat special (and less extreme than the isolated millisecond progenitors) or whether the scenario is simply wrong. Recent work suggested that evolution in a triple system might provide a viable formation channel to explain a small fraction of the isolated MSPs and peculiar binary pulsars \citep{port11a,frei11a}. Nonetheless, the ablation mechanism still appears as the most plausible general scenario for creating isolated MSPs.

Black widows and redbacks usually display extended eclipses at radio wavelengths, which are accompanied by rapid variations of the dispersion measure \citep{stap96b}. At X-ray energies, persistent emission is often visible in addition to pulsations, which likely results from the shocked relativistic wind colliding with the companion \citep{stap03a,tava93a}. Black widows and redbacks were thought to be a relatively small fraction of the total MSP population, but it has become clear that previous wide-field pulsar surveys have missed many sources (e.g. because of eclipses). In contrast, since energetic MSPs are efficient $\gamma$-ray emitters, targeted radio searches of $\gamma$-ray sources are biased towards finding them, especially since repeated searches of the same source have a better chance of catching it out of eclipse. About one third of the pulsars discovered by targeting $\gamma$-ray sources have been black widows or redbacks \citep{ray12b}. In the light of these discoveries, black widows might after all offer a viable channel for the formation of at least some of the isolated millisecond pulsars.

Optical and near-infrared observations of black widows and redbacks are an important probe of the state of the companion and the energetics of the system \citep[see, e.g.,][]{fruc88a,van-88b,stap01b}. The optical light is dominated by the companion -- the pulsar contributes a negligible amount -- and shows significant flux and color variations. Mostly, these reflect strong irradiation of the hemisphere facing the pulsar, but superposed on this are ellipsoidal variations due to the tidal distortion. Combined, these allow one to constrain the inclination and other physical parameters \citep{stap01b,reyn07a}. Combined with phase-resolved spectroscopy, this can be used to determine the component masses \citep[e.g.][]{van-11a,roma11a,roma12b}. Using this technique, \citet{van-11a} inferred that PSR~B1957+20 is likely very massive, $\sim\!2.40\pm0.12\,M_\odot$. Similarly, \citet{roma12b} find evidence that PSR~J1311$-$3430 is also heavyweight -- perhaps as much as $\sim\!2.7\,M_\odot$. These large masses suggest mass transfer was relatively effective, in contrast to what is inferred for other pulsar binaries \citep[e.g.,][]{lin11a,anto12a}, posing both an interesting quandary as to why this might be the case, and an opportunity to probe the upper mass limit of neutron stars.

While this kind of light curve modelling is also possible in other types of binary neutron star systems \citep[see, e.g.,][for low-mass X-ray binaries]{muno09b,wang11a}, black widows and redbacks are much cleaner systems. Indeed, the only source of optical light is the companion since there is no accretion disk or jet\footnote{PSR~J1023+0038 was observed to have a disk before but it has since disappeared \citep[see, e.g.,][]{arch09a}.}. Also, the irradiation is due to relativistic photons/particles which penetrate several optical depths inside the companion's photosphere. As a result, the atmosphere remains in quasi equilibrium and this avoids the formation of prominent emission line features. Finally, because the neutron star is a pulsar, the radio timing provides a set of accurate orbital parameters.

We have searched for the optical counterpart to four of the new black widow/redback systems, using {\em Gemini} data complemented by {\em New Technology Telescope} (NTT) and archival {\em Swift} data. In this paper, we present the optical discovery of these companions (Section~\ref{s:data}) and use their light curves to constrain the system parameters by modelling them using the synthesis code {\tt Icarus} (Section~\ref{s:results}). We find that the temperature increase on the dayside of the irradiated pulsar companions typically corresponds to a conversion efficiency $\sim\!15\%$ of the incident energy available from the host pulsars' rotational spin-down (Section~\ref{s:discussion}).

\section{Observations and Analysis}\label{s:data}

We were awarded 7.67\,hr on {\em Gemini North} (program GN-2010B-Q-77) to search for the counterparts of four energetic binary MSPs with the GMOS-N instrument \citep{hook04a}: PSRs~J1810+1744, J0023+0923, J2215+5135 and J2256$-$1024. The first three pulsars were found in a targeted GBT search of unidentified {\em Fermi} point sources (Bangale et al. in prep; Hessels et al. in prep; \citealt{hess11a}) while the last one was discovered during the 350\,MHz GBT pulsar drift scan survey (\citealt{boyl12a}; \citealt{lync12a}; Stairs et al. in prep). Table~\ref{t:parameters} presents the main properties of the targets, and a detailed discussion about each source follows in Section~\ref{s:discussion}. Our observing program had two main goals: detect the pulsar companions and, if they are detected, identify variability at the orbital period. Our observing strategy consisted of collecting data at four different epochs, with each observing session consisting of a 320-s $i$-band, 620-s $g$-band and 320-s $i$-band exposure sequence. Since we observed under non-photometric conditions (cloud cover: 90\%, image quality: 85\%, sky background: 80\%, water vapor: any), resolution was not an issue and we binned the EEV CCD detector by a factor $2 \times 2$ in order to reduce the readout time to 35\,s per exposure (fast mode) while still properly sampling the point spread function. Using this strategy, we were expecting to find inter-epoch variability and, given the short orbital periods of these binaries, maybe even intra-epoch variations (for the $i$-band).

\begin{table*}\footnotesize
\begin{center}
\caption{Measured and Inferred Source Parameters}
\begin{tabular}{lcccccc}
\hline
\hline
Quantity & J0023+0923\tablenotemark{a} & J2256$-$1024\tablenotemark{b} & J1810+1744\tablenotemark{a} & J2215+5135\tablenotemark{a} & B1957+20\tablenotemark{c} & J1023+0038\tablenotemark{c} \\
\hline
\multicolumn{7}{c}{Measured} \\
\hline
Right Ascension\tablenotemark{d}\dotfill & $00^h23^m16^s.89(2)$ & $22^h56^m56^s.39(1)$ & $18^h10^m37^s.28(1)$ & $22^h15^m32^s.68(1)$ & $19^h59^m36^s.77$ & $10^h23^m47^s.69$ \\
Declination\tablenotemark{d}\dotfill & $09^\circ23^\prime24^{\prime\prime}.18(20)$ & $-10^\circ24^\prime34^{\prime\prime}.37(12)$ & $17^\circ44^\prime37^{\prime\prime}.38(7)$ & $51^\circ35^\prime36^{\prime\prime}.45(10)$ & $20^\circ48^\prime15^{\prime\prime}.12$ & $00^\circ38^\prime41^{\prime\prime}.15$ \\
Gal. Longitude (degree) \dotfill & 111.38 & 59.23 & 44.64 & 99.87 & 59.20 & 243.49 \\
Gal. Latitude (degree) \dotfill & $-$52.85 & $-$58.29 & 16.81 & $-$4.16 & $-$4.70 & 45.78 \\
$f_{\gamma,\rm0.1-100\,GeV}$ \dotfill & $10.7 \pm 1.5$ & $10.1 \pm 1.4$ & $25.5 \pm 2.1$ & $10.9 \pm 1.6$ & $16.7 \pm 1.9$ & $5.4 \pm 1.0$ \\
($10^{-12}$\,erg\,s$^{-1}$\,cm$^{-2}$) \dotfill &  &  &  &  &  &  \\
$P_{\rm spin}$~(ms) \dotfill & 3.1 & 2.3 & 1.7 & 2.6 & 1.61 & 1.69 \\
$P_{\rm orb}$~(h) \dotfill & 3.3312 & 5.1092 & 3.5561 & 4.1401 & 9.1672 & 4.7543 \\
$T_{\rm asc. node}$~(MJD (TDB)) \dotfill & 55186.11343 & 54853.22391 & 55130.04813 & 55186.16449 & 48196.06352 & 54801.97065 \\
$x$~(lt-s) \dotfill & 0.035 & 0.083 & 0.095 & 0.47 & 0.089 & 0.343 \\
$DM$~(pc\,cm$^{-3}$) \dotfill & 14 & 14 & 40 & 69 & 29.12 & 14.33 \\
$A_V$\tablenotemark{e} \dotfill & 0.37 & 0.14 & 0.43 & 1.15 & \nodata & \nodata \\
\hline
\multicolumn{7}{c}{Inferred} \\
\hline
$d_{\rm DM}$~(kpc)\tablenotemark{f} \dotfill & 0.69 & 0.65 & 2.00 & 3.01 & 2.49 & 0.62 \\
$M_{\rm c}^{\rm min}$~($M_\odot$)\tablenotemark{g} \dotfill & 0.017 & 0.030 & 0.045 & 0.213 & 0.022 & 0.138 \\
$L_{\rm sd,\dot P}$~($10^{34}$\,erg\,s$^{-1}$)\tablenotemark{h} \dotfill & 1.51 & 3.95 & 3.97 & 5.29 & 16.0 & 9.82 \\
$L_{\rm \gamma}$~($10^{33}$\,erg\,s$^{-1}$)\tablenotemark{i} \dotfill & 1.28 & 1.21 & 3.05 & 1.30 & 2.00 & 0.65 \\
$a$~($R_\odot$)\tablenotemark{j} \dotfill & 1.27 & 1.69 & 1.33 & 1.53 & 2.49 & 1.65 \\
$R_L$~($R_\odot$)\tablenotemark{k} \dotfill & 0.13 & 0.22 & 0.19 & 0.36 & 0.29 & 0.34 \\
\hline
\multicolumn{7}{c}{Light Curves (minimum / maximum / quadrature)\tablenotemark{l}} \\
\hline
$i$-band \dotfill  & 24.3 / 21.7 / 22.7 & 24.3 / 20.8 / 22.1 & 22.3 / 19.5 / 20.3 & 19.5 / 18.6 / 18.9 & \nodata & \nodata \\
$g$-band \dotfill & 28.0 / 23.4 / 25.0 & 28.0 / 22.2 / 26.8 & 23.2 / 19.2 / 20.2 & 19.9 / 18.1 / 18.7 & \nodata & \nodata \\
\hline
\multicolumn{7}{c}{Light Curve Fitting} \\
\hline
inclination (degree) \dotfill & $58 \pm 14$ & $68 \pm 11$ & $48 \pm 7$ & $66 \pm 16$ & \nodata & \nodata \\
filling factor \dotfill & $0.30 \pm 0.30$\tablenotemark{m} & $0.40 \pm 0.20$ & $0.80 \pm 0.30$ & $0.99 \pm 0.03$ & \nodata & \nodata \\
$T_{\rm night}$~(K)\dotfill & $2900 \pm 700 $ & $2450 \pm 350$ & $\sim 4600\tablenotemark{n}$ & $4800 \pm 450$ & $\sim 2500$ & $\sim 5600$ \\
$T_{\rm day}$~(K)\dotfill & $4800 \pm 2000$ & $4200 \pm 700$ & $\gtrsim 8000\tablenotemark{n}$ & $6200 \pm 500$ & $\sim 5800$ & $\sim 6650$ \\
$T_{\rm irr}$~(K)\dotfill & $4600$ & $4100$ & $7800\tablenotemark{n}$ & $5550$ & $5750$ & $5580$ \\
$\epsilon_{\rm irr}$\dotfill & 0.17 & 0.07 & 0.60\tablenotemark{n} & 0.15 & 0.15 & 0.09 \\
\end{tabular}\label{t:parameters}
\tablenotetext{1}{Timing data ($P_{\rm spin}$, $P_{\rm orb}$, $x$, $DM$ and $L_{\rm sd,\dot P}$  from Hessels et al. (in prep.).}
\tablenotetext{2}{Timing data ($P_{\rm spin}$, $P_{\rm orb}$, $x$, $DM$ and $L_{\rm sd,\dot P}$ from Stairs et al. (in prep.).}
\tablenotetext{3}{PSRs~B1957+20 and J1023+0038 are shown for comparison. Data from the ATNF Pulsar Catalogue (\citealt{manc05a}; \url{http://www.atnf.csiro.au/research/pulsar/psrcat}).}
\tablenotetext{4}{Positions derived from optical astrometry. Uncertainties are dominated by the catalog systematics of 0\farcs07 in UCAC3 (J0023+0923, J1810+1744 and J2215+5135) and 0\farcs1 in SDSS DR7 (J2256$-$1024).}
\tablenotetext{5}{Total reddening along the line of sight (see Section~\ref{s:results}).}
\tablenotetext{6}{Distance based on the dispersion measure NE2001 model \citep{cord02a}.}
\tablenotetext{7}{Minimum companion mass assuming a 1.4\,$M_\odot$ pulsar and a 90$^\circ$ orbital inclination.}
\tablenotetext{8}{Spin-down luminosity inferred from $P_{\rm spin}$ and its derivative, and a moment of inertia of $10^{45}{\rm\,g\,cm^2}$.}
\tablenotetext{9}{$\gamma$-ray luminosity estimated from the $\gamma$-ray flux and dispersion-measure distance.}
\tablenotetext{10}{Orbital separation $a = x (1+1.4/M_{\rm c}^{\rm min})$.}
\tablenotetext{11}{Companion roche radius $R_L = 0.46 a [M_{\rm c}^{\rm min}/(M_{\rm c}^{\rm min}+1.4)]^{1/3}$.}
\tablenotetext{12}{The reported magnitudes are inferred from the light curve modelling presented in the text.}
\tablenotetext{13}{The probability distribution for this parameter is highly non-Gaussian, hence the value must be taken with caution. The median value is 0.40, the mode at $\sim 0.15$ and the distribution extends with a heavy tail all the way to unity.}
\tablenotetext{14}{As explained in Section~\ref{s:j1810}, we believe that our analytic estimate is more robust than the numerical values for the case of PSR~J1810+17. Hence we present the analytic results in this table.}
\end{center}
\end{table*}

We reduced the data following standard procedures, implemented using custom Python scripts. We used standard {\em Gemini} nightly calibration data to remove the bias and flat-field our science frames. We registered our frames astrometrically relative to the UCAC3 catalog for PSRs J1810+1744, J0023+0923 and J2215+5135, and relative to SDSS for PSR~J2256$-$1024 (the only source that falls within the coverage of SDSS Data Release 7). For all but PSR~J0023+0923, a few tens of reference stars fall within the central frame of the CCD detector and hence our calibration yielded positional uncertainties dominated by systematic uncertainties, of 0\farcs07 and 0\farcs1 in UCAC3 and SDSS DR7, respectively. For PSR~J0023+0923, only three UCAC3 stars fall on the central CCD chip. In this case, we calibrated the image that looked the cleanest using the UCAC3 catalogue, and then calibrated the other images in the same band relative to the reference one using a list of bright stars found in all images. The measured optical positions are reported in Table~\ref{t:parameters} and agree with the radio position derived from the timing.

We performed aperture photometry using an extraction radius of 5 pixels (i.e. 0.73\,arcsec at the plate scale of 0.146 arcsec per binned pixel), and sky inner and outer annuli of 10 and 15 pixels, respectively. We calibrated our photometry against bright, non-saturated stars appearing in the SDSS DR7 catalogue for PSR~J2256$-$1024, and USNO-B1 in the case of the three other targets. For USNO-B1, we converted the catalogue magnitudes from the photographic B, R, and I magnitudes to the {\it ugriz} system using the transformation of \citet{jord06a}, $i-I = (0.247 \pm 0.003) (R-I) + (0.329 \pm 0.002)$, for the $i$-band, and from \citet{lupt05a}, $B-g = 0.3130 (g-r) + 0.2271$ and $R-r = -0.1837 (g-r) - 0.0971$, for the $g$-band. In the case of PSR~J1810+1744, there is a neighbouring star that might contaminate our aperture photometry. For this reason, we performed point-spread function photometry using a Moffat profile $f(r) \propto (1 + r^2/\sigma^2)^{-\beta}$, with $\beta = 3$ and $\sigma$ optimized using a set of bright, non-saturated stars for each frame. Reference stars were fitted individually while the immediate vicinity of PSR~J1810+1744 was simultaneously fitted for the source and the two other stars located East and North-East from it.

Our complete photometric results are available online in Tables~\ref{t:j0023}, \ref{t:j2256}, \ref{t:j1810} and \ref{t:j2215}. Our photometric errors were calculated by adding in quadrature the sky background, the photon counting noise and the intra-band relative zero-point. The zero-point calibration errors correspond to the standard deviation of the mean of the zero-point calibration for each band to the reference catalog stars, which were added in quadrature to the catalog systematic calibration. In the case of PSR~J0023+0923, very few catalog stars overlap with our field and hence the band calibration is poorer than for the other systems analyzed here. The SDSS systematic calibration error is 0.02\,mag (for PSR~J2256$-$1024), while that of USNO-B1.0 is 0.3\,mag (for PSRs~J1810+1744, J0023+0923 and J2215+5135).

The field of PSR~J2215+5135 was also serendipitously observed by the UVOT instrument on-board {\em Swift} while it was monitoring a nearby gamma-ray burst. We performed photometric reduction of the publicly available data\footnote{\url{http://heasarc.gsfc.nasa.gov}}, which were all obtained in the uvw1-band (\citealt{pool08a}; $\lambda_{\rm c} = 260$\,nm, $\Delta\lambda = 69.3$\,nm). For the UVOT data, we used a 5 pixel aperture (i.e. 5\,arcsec at 1.004\,arcsec\,pixel$^{-1}$) and sky inner and outer annuli of 15 and 30 pixels, respectively. We took the photometric zero-point from \citet{bree11a}, ${\rm zp}_{\rm uvw1} = 18.95 \pm 0.03$ (AB system), to convert our count rate to magnitude. Note that the aperture correction is negligible for our aperture size\footnote{\url{http://heasarc.gsfc.nasa.gov/docs/swift/analysis/uvot\_digest/apercor.html}}.

We also obtained some exploratory exposures during an observing run at the NTT using the tri-band ULTRACAM imager \citep{dhil07a} and managed to obtain an additional $z$-band and $g$-band image for PSR~J2256$-$1024. The data processing was performed using the ULTRACAM data reduction pipeline\footnote{\url{http://deneb.astro.warwick.ac.uk/phsaap/software/}}.

\section{Results}\label{s:results}

\begin{figure*}
\centering
\includegraphics[height=6cm]{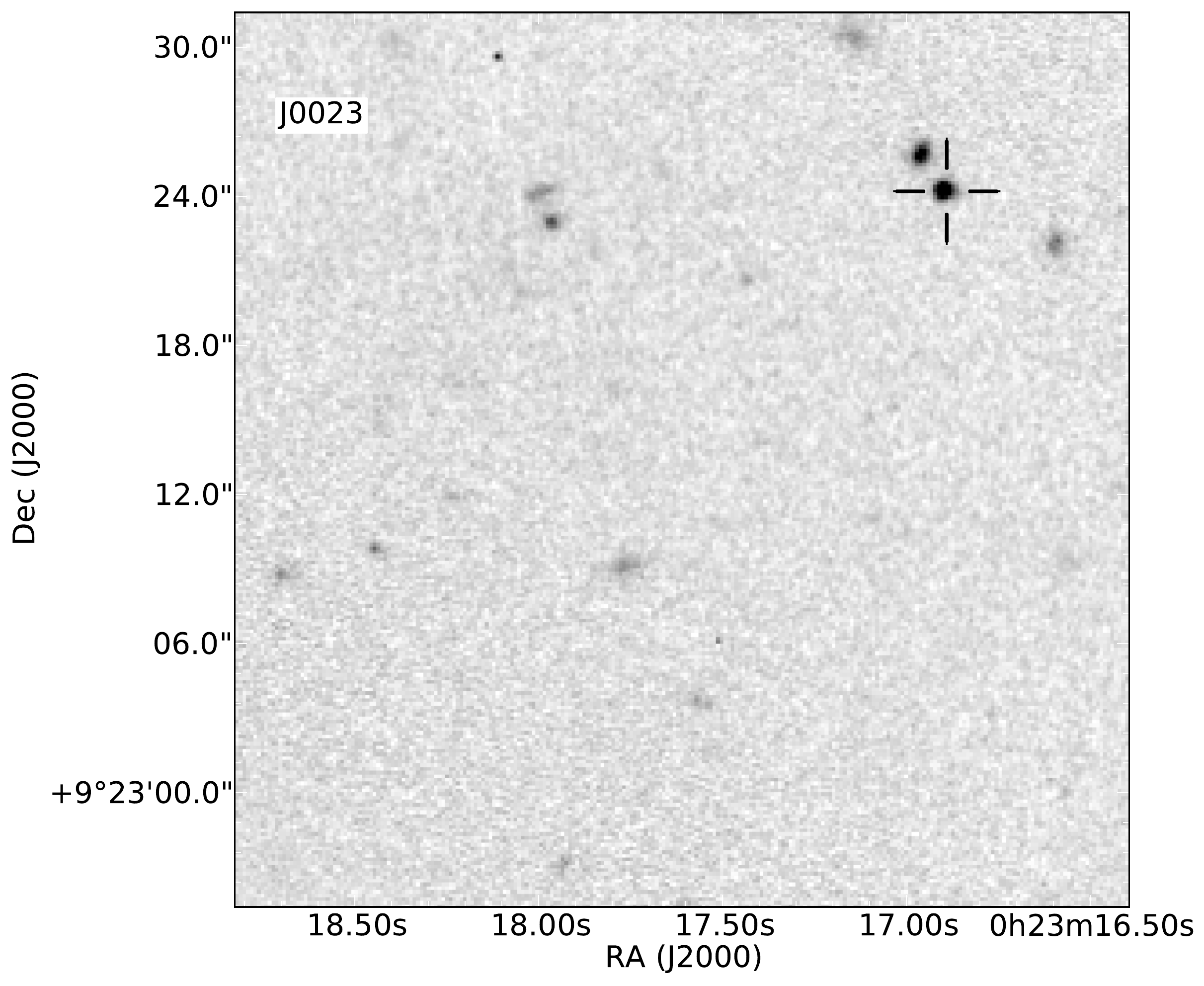}
\includegraphics[height=6cm]{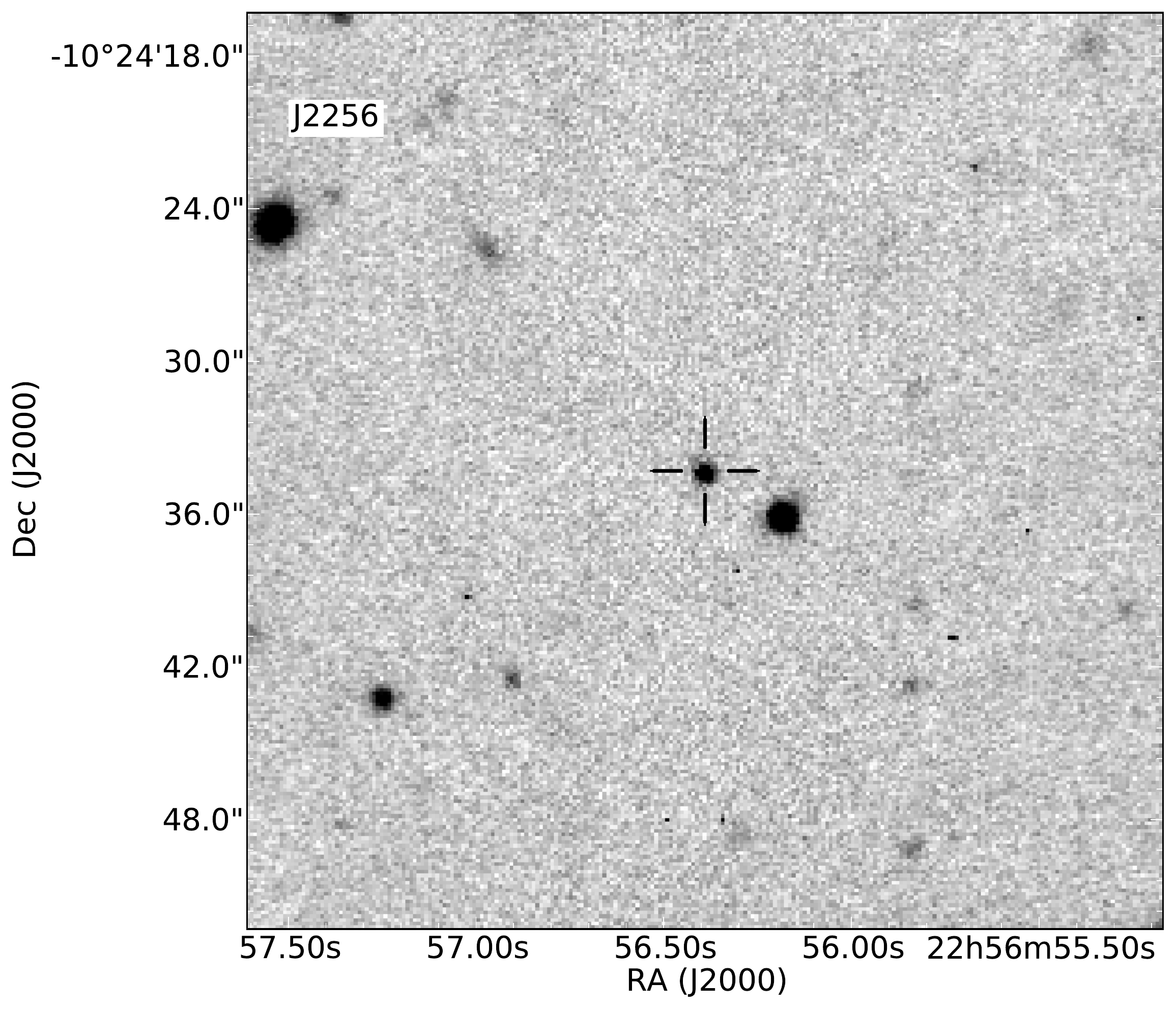}\\
\includegraphics[height=6cm]{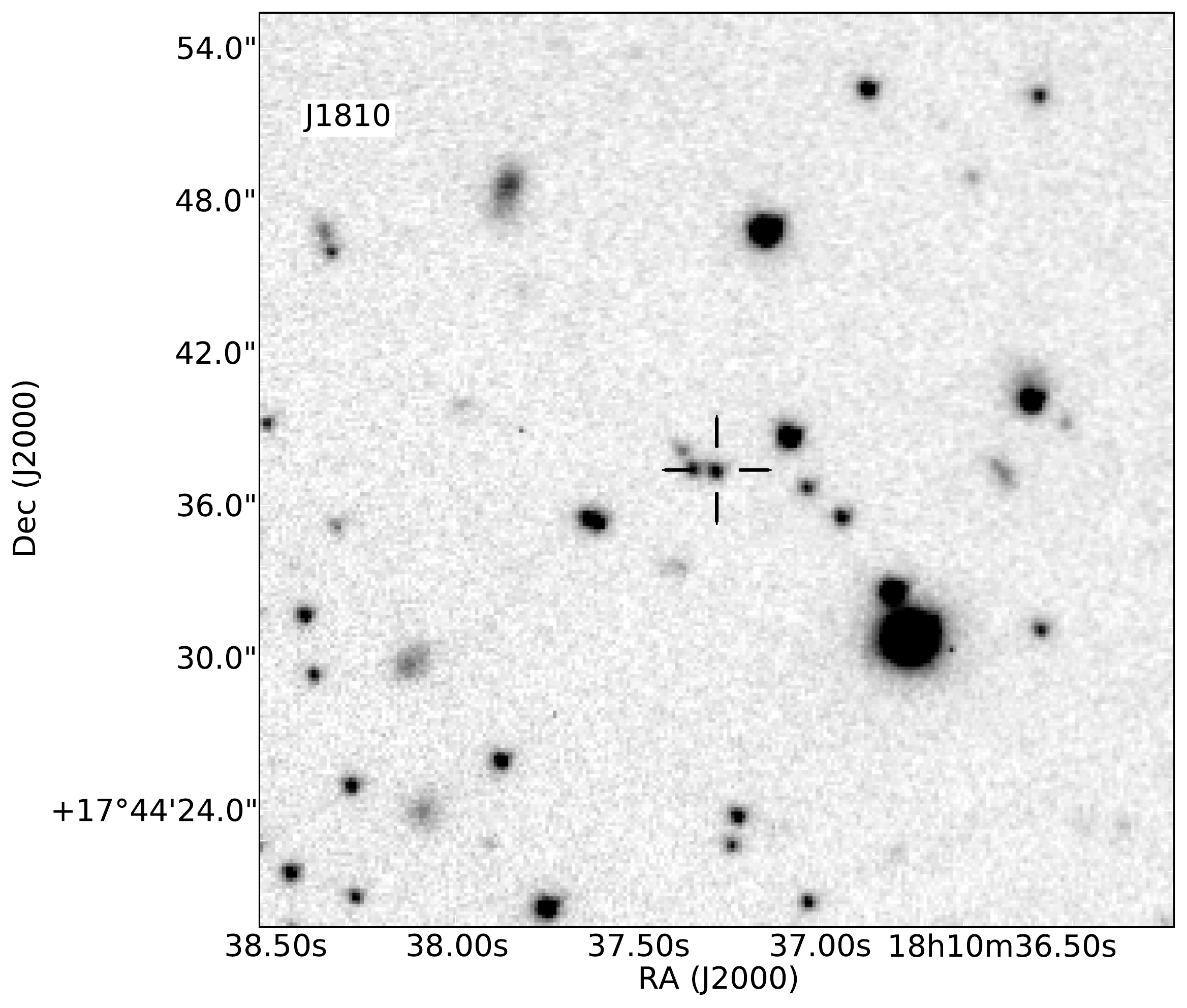}
\includegraphics[height=6cm]{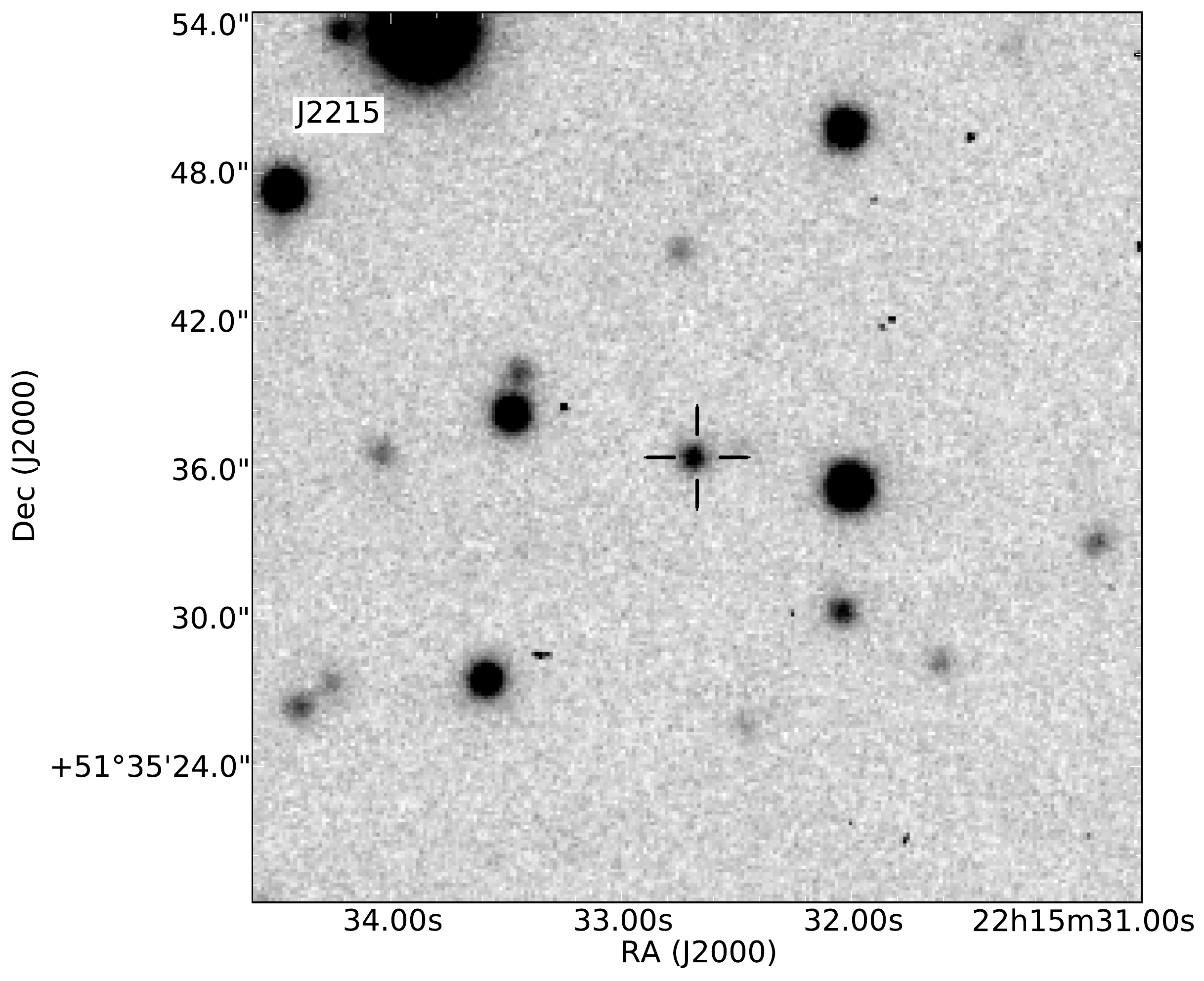}
\caption{Finding chart of the four irradiated pulsar companions presented in this paper, observed in $i$-band with GMOS-North. In each panel, the location of the counterpart is indicated with crosshairs.\label{f:finding_charts}}
\end{figure*}

\begin{figure}
\centerline{\hfill
\includegraphics[width=0.99\hsize]{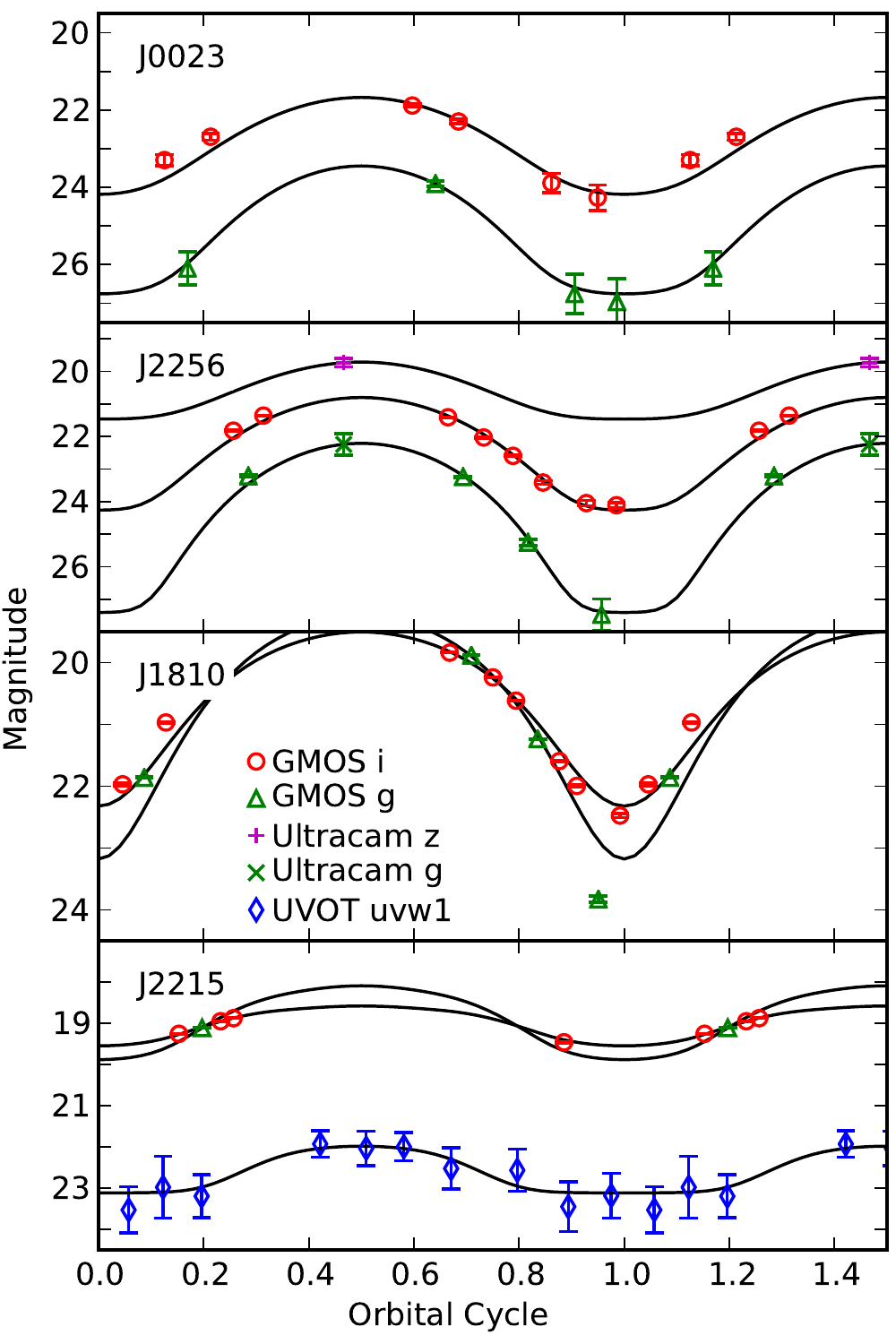}%
\hfill}
\caption{Light curves of the four irradiated pulsar companions. GMOS $i$, GMOS $g$, ULTRACAM $z$, ULTRACAM $g$ and UVOT $uvw1$ bands are marked by red circles, green triangles, magenta plus-signs, green cross-signs, and blue diamonds, respectively. The black lines display the best-fit light curve calculated using {\tt Icarus}.\label{f:lightcurves}}
\end{figure}

We found optical counterparts to all four pulsar binaries (see Figure~\ref{f:finding_charts}). The association of the optical counterparts with the irradiated pulsar companions was confirmed in all cases by variability at the known orbital periods, as can be seen in the light curves shown in Figure~\ref{f:lightcurves} (see also online Tables~\ref{t:j0023}, \ref{t:j2256}, \ref{t:j1810} and \ref{t:j2215}).

As discussed in Section~\ref{s:introduction}, the light curves can be used to constrain several parameters of the system. The orbital modulation is caused by a combination of irradiation, which produces a maximum at the superior conjunction of the companion, and ellipsoidal variations, which yield maxima at quadrature. From the colors at superior and inferior conjunction of the companion, one can constrain the dayside and nightside temperatures and thus irradiation. For poorly sampled data like ours, the nightside temperature cannot be constrained very well, since it is degenerate with the orbital inclination. Good coverage is required to distinguish between the high-inclination case of seeing only the brighter, unirradiated nightside at inferior conjunction of the companion, or the low-inclination case where one observes a dimmer nightside combined with a sliver of the irradiated side. However, for strongly irradiated systems, where the dayside is much hotter than the nightside, the irradiated temperature can be determined relatively securely from just the colors near superior conjunction of the companion. The amplitude of ellipsoidal variations directly depends on the filling factor of the companion. Since the amplitude is small relative to irradiation, one must possess good quality sampling near quadrature in order to constrain the filling factor accurately. However, the physical size of the star is also directly related to its apparent luminosity given its distance, and thus any independent information on the latter, such as from the dispersion measure, can be used to constrain the filling factor. For small filling factors,\footnote{We define the filling factor as the ratio of the volume-averaged stellar radius and the volume-averaged Roche-lobe radius.} below $\sim\!0.55$, the scaling between filling factor and distance is linear (as expected), but for larger filling factors its dependence becomes sub-linear, as the ellipsoidal variations and gravity darkening start to counteract the irradiation at maximum light (superior conjunction), leading to a $\sim\!20\%$ reduction relative to the linear extrapolation when the companion fills its Roche lobe. Finally, we note that constraining parameters such as inclination and filling factor is generally limited by the quality of the sampling and the uncertainty on individual data points, whereas the determination of the temperatures are mostly affected by the absolute calibration of each band.

To get first-order estimates of the system parameters, we modelled the light curves of these irradiated binaries using {\tt Icarus}\footnote{\url{https://github.com/bretonr/Icarus}} \citep{bret11a} along with BTSettl atmosphere models \citep{alla03a,alla07a,alla10a} available through the Phoenix web simulator\footnote{\url{http://phoenix.ens-lyon.fr/simulator/}}. The free parameters in the model are the mass ratio~$q$, the orbital inclination~$i$, the distance~$d$ and reddening~$A_V$ to the system, the companion's filling factor~$f=R/R_{\rm L}$ (where $R$ is the volume-averaged radius and $R_{\rm L}$ the volume-averaged radius of the Roche lobe), and its dayside and nightside temperatures $T_{\rm day}$ and $T_{\rm night}$. We assumed that the companion is tidally locked to the pulsar and added Gaussian priors on the distance and reddening. We set the mean of the distance priors to the values inferred from dispersion measure using the NE2001 model \citep{cord02a}, and, since our sources are well outside the dust layer, we estimated the reddening as being equal to the total along each line of sight as inferred by \citet{schl98a} (see values in Table~\ref{t:parameters}). In both cases, we took standard deviations corresponding to 25\% relative uncertainties. Also, since the mass ratio is unconstrained without spectroscopic information, we imposed that the mass of the pulsar lies in the range $1-3\,M_{\odot}$. For the model fitting, we used the uncertainties from Tables~\ref{t:j0023}, \ref{t:j2256}, \ref{t:j1810} and \ref{t:j2215} on individual data point, and a band-to-band uncertainty corresponding to our band calibration errors to which we added in quadrature an additional 0.1\,mag to account for systematics in the zero-point calculation inherent to the atmosphere models (this is important only for PSR~J2256$-$1024 for which the photometry could be accurately tied to SDSS). In the following subsections, we discuss the main results of the light curve modelling (see Figure~\ref{f:lightcurves} for best-fit light curves).

\subsection{PSR~J0023+0923}

PSR~J0023+0923 stands out from other black widows in its relative long spin period, and thus relatively low spin-down luminosity, which makes the expected irradiation effects on the companion relatively mild. The data constrain the colors at maximum and minimum light well, with $g-i\simeq1.5$ and $3.0$, respectively. These correspond to flux-averaged temperatures of $\sim\!4500$ and $3200\,$K, respectively, close to what we infer from our models: $4800 \pm 2000$ and $2900 \pm 700$\,K, respectively. Note that the latter dayside temperature is strictly for the part directly facing the pulsar, hence quite a bit higher than the flux-averaged temperature inferred from the color. The substantial change in color shows that much of the irradiated side must be hidden at inferior conjunction of the companion, and our models demonstrate that this excludes orbital inclinations lower than $i\lesssim40^\circ$; the preferred value is $i=58 \pm 14^\circ$. 

The filling factor appears relatively small at $0.30 \pm 0.30$, though one must be cautious since the probability distribution is highly non-Gaussian with a mode at $\sim 0.15$ and an extended heavy tail all the way to a filling factor of unity (median value at 0.40). This could imply a much smaller companion than inferred for the prototype black widow, PSR~B1957+20, which has a filling factor fairly close to unity \citep{reyn07a}. Indeed, the nominal implied size, of $\sim\!0.05\,R_\odot$ is smaller than is possible for a solar-composition object, suggesting that the filling factor is on the upper end of our inferred range (and thus the distance somewhat larger than inferred from the dispersion measure; see Section~\ref{s:discussion}). However, indirect evidence that the companion is not close to filling its Roche lobe comes from the fact that this source shows no radio eclipses at 350\,MHz\footnote{Radio eclipses are strongly radio-frequency dependent, being more dramatic at lower frequencies \citep[see, e.g.][]{arch09a, stap96b}.}, unlike most black widows (Hessels et al., in prep.; though we could also be seeing the system relatively face-on).

\subsection{PSR~J2256$-$1024}

This system appears to be quite similar to PSR~J0023+0923, showing slightly redder colors. Near maximum, the fortuitously phased ULTRACAM points yield $g-z\simeq2.5$, while near minimum, the Gemini data yield $g-i\simeq3.5$, corresponding to flux-averaged temperatures of $\sim\!3700$ and $3100\,$K, respectively. Since for PSR~J2256$-$1024 our photometric calibration is more secure, our fits yield well-constrained day and nightside temperatures, of $4200 \pm 700$ and $2450 \pm 350$\,K, respectively. The modelling favors an intermediate orbital inclination, $i=68 \pm 11^\circ$.

Like for PSR~J0023+0923, the best-fit filling factor is small, $0.40 \pm 0.20$, although since the orbit is somewhat wider, the inferred radius of $0.09\,R_\odot$ is not inconsistent with a solar-composition, degenerate object. Contrary to PSR~J0023+0923, however, this system shows radio eclipses, suggesting that, perhaps, the size is underestimated because the distance derived from the dispersion measure is too small (see also Section~\ref{s:discussion}), and/or that it is simply closer to being observed edge-on. The latter explanation appears slightly unlikely though given that our lightcurve fitting yield similar orbital inclination ranges (see Table~\ref{t:parameters}).

\subsection{PSR~J1810+1744}\label{s:j1810}

Like PSRs~J0023+0923 and J2256$-$1024, this system is a canonical black widow. One should expect a larger irradiation than in the case of PSRs~J0023+0923 and J2256$-$1024, given the combination of a compact 3.6-hr orbit and a more energetic pulsar (in fact, this is the fourth fastest-spinning pulsar known in the Galactic field). Its light curves are quite puzzling since we did not find a combination of parameters that successfully account for all the data points. Indeed, the $i$-band light curve in particular seems inconsistent with being symmetric around minimum light, being dimmer before minimum than after. Asymmetric light curves have also been seen for PSR~J2051$-$0827 \citep{stap01b}, though in that case the source was brighter before minimum than it was afterwards, and the light curve showed variations between different sets of observations (in contrast to what is seen for PSR~B1957+20; e.g., \citealt{reyn07a}). In the $g$-band light curve, the data point near minimum is significantly offset from the predicted value. As we mentioned in Section~\ref{s:data}, particular care was taken for the flux extraction of this source. Hence we believe that the pulsar companion was intrinsically faint at this orbital phase. We note that some irradiated pulsar companions, such as PSR~J1311$-$3430 \citep{roma12b}, display flares so it is not impossible that PSR~J1810+1744 also shows variability. Further observations with better orbital coverage should help understand the odd behavior of this source.

As a result of the poor fit, we can only give qualitative constraints on the parameters. Given the large modulation, the inclination has to be relatively large, though the lack of flattening near minimum implies that the dayside has to remain partly visible; we find $i \simeq 48 \pm 7^\circ$. Our best-fit model also yields a very hot dayside, with $\sim\!10000\,$K, and a much colder nightside, with $\sim\!3100\pm850\,$K. Given the larger distance to the system, $\sim\!2$\,kpc, a larger filling factor can be accommodated. Our data are consistent with a Roche-lobe filling star, though the poor fit and the lack of photometric coverage near maximum yields large uncertainties, hence explaining a heavy tail in the distribution of possible filling factor at lower values. The inferred filling factor of $0.80 \pm 0.30$, which implies a companion radius of $0.15\,R_\odot$.

Given the poor fit, however, we caution that the above results cannot be taken at face value. In particular, while our data cover minimum and thus directly constrain the nightside temperature (for $g-i\simeq1.3$, one infers a flux-averaged temperature of $\sim\!4600\,$K), maximum is not covered and hence we have no direct constraints on the dayside temperature. Nevertheless, the dayside must be relatively hot, since the source has $g-i \simeq 0$ near quadrature, implying a flux-averaged temperature of $\sim\!7500\,$K. For many black widows, the color does not vary strongly between quadrature and maximum -- see PSRs~J0023+0923 and J2256+1024 above, as well as PSR~B1957+20 \citep{call95a}. Given a dayside color of $g-i\lesssim0$, and taking into account that the dayside temperature is somewhat hotter than the flux-averaged one, we infer a lower limit to the dayside temperature of $\gtrsim\!8000\,$K. We will use this limit below. Using these values also reduces the irradiation efficiency from the formal best-fit value of 1.5 to 0.60, which make it more plausible energetically.

\subsection{PSR~J2215+5135}

Unlike the systems we have already discussed, PSR~J2215+5135 has a significantly more massive companion and is best characterized as a redback, with orbital properties similar to the prototype, PSR~J1023+0038 \citep{arch09a}. Like PSR~J1023+0038, it shows only modest brightness variability. The Gemini data give only a single color, $g-i\simeq0.3$, implying a flux-averaged temperature of $\sim\!6600\,$K near quadrature. The presence of UVOT data proves to be valuable at constraining parameters of this system, with the caveat that ultraviolet data are sometimes contaminated by chromospheric emission if the companion is magnetically active. The system appears to be viewed at an intermediate orbital inclination ($\sim\!70^\circ$) and a dayside temperature of $6200 \pm 500$\,K. Like other redback companions, the nightside is much hotter ($\sim\!4800$\,K) than that of black widow companions. The filling factor is very tightly constrained to being Roche-lobe filling, which is consistent with the occurrence of radio eclipses for $\sim 50$\% of the orbit (Hessels et al., in prep).

\section{Discussion}\label{s:discussion}

As can be seen in Figure~\ref{f:lightcurves}, the model light curves generally agree well with our data. While our light curves are not sufficiently well-sampled to produce strong constraints on the system parameters, they allow us to address two interesting aspects: the efficiency of the irradiation and the extent to which the companion fills its Roche lobe.

To estimate the efficiency of the irradiation, i.e., the effective fraction $\epsilon_{\rm irr}$ of the spin-down luminosity incident on the companion that is absorbed and re-radiated, we assume that the irradiating flux is thermalized and reradiated locally, i.e., that it simply adds to the intrinsic flux wherever it impinges. Then, from the hottest point, one can derive a characteristic ``irradiation temperature'' $T_{\rm irr}^4 = T_{\rm day}^4 - T_{\rm night}^4$, which is related to the pulsar's spin-down luminosity, $L_{\rm sd}$, as $\epsilon_{\rm irr} L_{\rm sd} = 4\pi a^2 \sigma T_{\rm irr}^4$, where $a$ is the orbital separation and $\sigma$ is the Stefan-Boltzmann constant. Note that we implicitly assume that the pulsar spin-down energy is isotropically radiated.

In Table~\ref{t:parameters}, we list our inferred irradiation temperatures as well as the implied irradiation efficiencies. With the exception of PSR~J1810+1744, which we will discuss below, we find that the typical values of the irradiation efficiency of the new systems presented in this paper are consistent with those of PSRs~B1957+20 and J1023+0038. Moreover, the other known irradiated pulsar systems (PSRs~J2051$-$0827, J1311$-$3430, J2339$-$0533 \citealt{stap01b,roma12b,roma11a}) also display similar efficiencies ($\gtrsim\!30\%$, $\sim\!30\%$ and $\sim\!15\%$, respectively)\footnote{PSR~J2051$-$0827 has a somewhat large value efficiency. The efficiency of PSR~J2339$-$0533 is based on the radio timing from Ray et al. (in prep.).}. We conclude that the typical irradiation efficiency factor in these systems lies in the range $10-30$\%, with 15\% being a representative figure. Note that the intrinsic spread in values might be smaller if one compensates for the fact that the orbital separation is poorly known given the uncertainties in the inclination, the pulsar mass and its moment inertia.

The energetics derived from the irradiation of the companions are consistent with the idea that the relativistic wind, which is powered by the rotational spin down of the neutron star, is the major driver of the heating mechanism. The case of PSR~J1810+1744 is, however, puzzling. Its rather large inferred irradiation efficiency not only departs from the other known irradiated pulsar systems but it also implies an input energy larger than the nominal spin-down luminosity (though our analytic estimate of the lower limit of $\sim\!0.60$ is below unity). One cause may be that our assumption of an isotropic wind is not justified. For instance, if the pulsar is aligned with the orbit and emits its wind preferentially in the equatorial plane (as is the case for, e.g., the Crab and Vela pulsars), the irradiation efficiency would be reduced. From our Gemini data, there is a clear indication that both $g$ and $i$ band light curves are not symmetric, with the companion being brighter after its inferior conjunction (phases $\sim 0.0-0.15$) than before (phases $\sim 0.85-1.0$). This could be an indication of non-isotropic heating or heat redistribution at the surface of the star. More detailed light curves of this system would help resolve these issues.

Radio eclipses are observed in three out of four of these systems (Hessels et al. in prep.; Stairs et al. in prep.) and large increases in the dispersion measure at the ingress and egress in similar systems indicates total intra-binary electron column densities of $N_e \sim 10^{16}{\rm\,cm^{-2}}$ \citep[see, e.g.,][]{fruc88b}. Given that eclipses are coincident with the inferior conjunction of the companions, plasma must certainly be surrounding them and hence some form of mass loss from the companion is required. In the case of a nearly Roche-lobe filling star, material is loosely bound to the surface and can be peeled off easily when exposed to a relativistic pulsar wind. Our work suggests that, however, some of the irradiated pulsar companions that we have studied (PSRs~J0023+0923 and J2256$-$1024) are not close to filling their Roche lobe.

As mentioned in the previous section, a first possibility is that the distances inferred from the dispersion measures of these pulsars are underestimated. It is now well established that for pulsars located far off the Galactic plane, the measured parallactic distances tend to be larger than the DM distances \citep[see, e.g.,][]{gaen08a,robe11a}. \citet{robe11a} shows that based on 13 sources with measured parallax and Galactic latitude larger than $10^\circ$, $d_{\rm DM} / d_{\rm parallax} = 0.66 \pm 0.26$. Since the filling factor is correlated with the distance, we ran another set of fits using priors on the distance rescaled using the above conversion factor and error. As a result, we found that the typical filling factor for PSRs~J0023+0923, J1810+1744 and J2256$-$1024 did not change significantly\footnote{PSR~J2215+5135 is already constrained to be Roche-lobe filling.}. This comes from the fact that the dayside temperature are not very precisely constrained due to the large systematic uncertainties in the absolute calibration of the bands. Consequently, the larger distance priors tends to increase the dayside temperatures rather than changing the filling factors as one would expect. We also ran another set of fits, this time by holding the filling factor of the companions to unity and removing the distance priors in order to see how much further these systems would need to be located in order to match the observed fluxes. We found that the DM distances would need to be off by a factor 8.5, 1.2 and 2.3 for the three above sources, respectively.

While it is not excluded that our distance estimates are wrong, it appears unlikely that PSRs~J0023+0923 and J2256$-$1024's companions are Roche-lobe filling. Whether nearly Roche-lobe filling stars are required in order to explain the radio eclipses of the pulsars is uncertain since neither the mechanism supplying particles to the plasma nor the role of the pulsar at triggering it are understood. Caution should be taken before drawing definitive conclusions and precise distance constraints from parallactic measurements would help shed light on this. Better-quality multi-color light curves will improve the measurement of the orbital inclination and address the contribution of ellipsoidal variations, hence also help to constrain the filling factor. It is worth mentioning that the second black widow system to be found, PSR~J2051$-$0827, also displays puzzling light curves. Previous work highlighted an ambiguous behavior either indicating a filling factor near unity or closer to 50\% \citep{stap01b}.

A common feature of black widow and redback systems is the presence of non-secular orbital period derivatives in the radio timing \citep[see, e.g.][]{arzo94a,laza11a,arch09a}. It has been suggested that gravitational quadrupole coupling of the companion with the orbit might explain the orbital variability \citep{appl94a}. The dissipated tidal energy would drive convection, which would power a dynamo-induced magnetic field and provide a significant source luminosity. While such a mechanism would be compatible with the amplitude of the orbital variations of PSR~B1957+20 and the luminosity of its companion \citep{appl94a}, it would require some fine tuning -- namely a $\sim$50\% filling factor -- in order to work for PSR~J2051$-$0827 \citep{laza11a}. These newly discovered irradiated pulsar systems could therefore provide extended leverage to test the gravitational quadrupole theory, since the Roche-lobe under-filling systems are predicted to display smaller orbital variability.

\section{Conclusion}\label{s:conclusion}

Our view of the binary pulsar population is currently shifting toward a new paradigm. Until the launch of {\em Fermi}, the bulk ($\sim\!90\%$) of the known population in the Galactic field consisted of pulsar--white dwarf systems \citep{lori04a}, while the remaining pulsar binaries had neutron star, main sequence, very low-mass star or planet companions. Only about four of the known binary pulsars in the Galactic field were irradiated systems like those presented here. As of today, the number of irradiated systems has increased to over twenty members and candidates, which implies that they now account for about 10\% of the binary pulsar population outside of globular clusters (based on the ATNF catalog, \citealt{manc05a}). It is clear that a large selection bias against finding these binaries in classical radio surveys existed until high-energy missions were added in the picture --- and yet they still remain challenging to find. New radio pulsar surveys, benefiting from multibeam receivers and larger bandwidth, are also contributing to finding irradiated systems in blind searches, as was the case for PSR J2256$-$1024. Black widows and redbacks therefore constitute a fundamental component of the pulsar ecosystem, a component that dominates among the fastest-spinning MSPs \citep{hess08b}.

The work presented here shows that the spin-down luminosity of pulsars is a good indicator of the level of irradiation sustained by their companions. We found that these systems display a rather universal irradiation efficiency $\epsilon_{\rm irr} \sim 10-30\%$ for reprocessing the incoming energy flux from the pulsar's spin-down into heat on the companion's surface. As a result, one may easily estimate the brightness and amplitude of the optical light curves due to irradiation in these pulsar binaries, provided an orbital separation, if the pulsar spin-down luminosity is known from timing or, alternatively, from the $\gamma$-ray luminosity.

The typical reflection albedo of stars with temperatures in the range $2000-10000$\,K is between 0.5 and 1.0 \citep{clar01a} for atmospheres that are convective and in radiative equilibrium, respectively. Given the above 15\% irradiation efficiency, the above albedos imply that $10-30$\% of the energy from the spin-down luminosity would actually reach the companion. It is worth noting that the temperature of PSR~J1810+1744 suggests that its outer envelope might be radiative, as opposed to convective in the other systems presented here. If so, the observed irradiation efficiency would be a factor 2 larger because of the difference in the stellar albedo and this could partly explain why it appears unusual. If further studies of these systems find bow shock nebulae (like for PSR~B1957+20), it would allow an independent measurement of the energy loss by the pulsar, which would make for an interesting comparison with that inferred from the irradiated companion \citep[see, e.g.,][]{van-08a}.

The possibility that some of these pulsar companions do not fill their Roche lobe leads one to ponder the underlying cause for radio eclipses in these systems. What is the mechanism responsible for replenishing the plasma responsible for the eclipses? Other missing pieces of the evolutionary puzzle are: Were these companion stars closer to filling their Roche lobe in the past or have they contracted thermally since the mass transfer episode has terminated? Can black widows and redbacks completely destroy their companion and eventually become the isolated millisecond pulsars that we observe? Or, are we simply observing the systems that are incapable of destroying their companions on a relatively short timescale?

The fact that some of these irradiated systems are relatively bright also offers an interesting opportunity for spectroscopic follow-up. Radial velocity curves should allow for the measurement of the component masses, and test whether these neutron stars typically are more massive, as found for the single system studied so far. Further, detailed photometry work will also allow one to investigate the possible asymmetry in the light curves of PSR~J1810+1744 and the reason for its anomalously large dayside temperature.

\acknowledgments
{JWTH is a Veni Fellow of the Netherlands Foundation for Scientific Research (NWO). Pulsar research at UBC and UofT is supported by NSERC Discovery Grants. This work was partially supported by the NASA Fermi Guest Observer Program. Based on observations obtained at the Gemini Observatory, which is operated by the Association of Universities for Research in Astronomy, Inc., under a cooperative agreement with the NSF on behalf of the Gemini partnership: the National Science Foundation (United States), the Science and Technology Facilities Council (United Kingdom), the National Research Council (Canada), CONICYT (Chile), the Australian Research Council (Australia), Minist\'{e}rio da Ci\^{e}ncia, Tecnologia e Inova\c{c}\~{a}o (Brazil) and Ministerio de Ciencia, Tecnolog\'{i}a e Innovaci\'{o}n Productiva  (Argentina).}

{\it Facilities:} \facility{{\em Fermi}}, \facility{{\em GBT}}, \facility{{\em Gemini}}, \facility{{\em NTT}}, \facility{{\em Swift}}.

\bibliography{blackwidow_letter}

\begin{thebibliography}{53}
\expandafter\ifx\csname natexlab\endcsname\relax\def\natexlab#1{#1}\fi

\bibitem[{{Allard} {et~al.}(2007){Allard}, {Allard}, {Homeier}, {Kielkopf},
  {McCaughrean}, \& {Spiegelman}}]{alla07a}
{Allard}, F., {Allard}, N.~F., {Homeier}, D., {Kielkopf}, J., {McCaughrean},
  M.~J., \& {Spiegelman}, F. 2007, \aap, 474, L21

\bibitem[{{Allard} {et~al.}(2003){Allard}, {Guillot}, {Ludwig}, {Hauschildt},
  {Schweitzer}, {Alexander}, \& {Ferguson}}]{alla03a}
{Allard}, F., {Guillot}, T., {Ludwig}, H.-G., {Hauschildt}, P.~H.,
  {Schweitzer}, A., {Alexander}, D.~R., \& {Ferguson}, J.~W. 2003, in IAU
  Symposium, Vol. 211, Brown Dwarfs, ed. {E.~Mart{\'{\i}}n}, 325

\bibitem[{{Allard} {et~al.}(2010){Allard}, {Homeier}, \& {Freytag}}]{alla10a}
{Allard}, F., {Homeier}, D., \& {Freytag}, B. 2010, ArXiv e-prints

\bibitem[{{Antoniadis} {et~al.}(2012){Antoniadis}, {van Kerkwijk}, {Koester},
  {Freire}, {Wex}, {Tauris}, {Kramer}, \& {Bassa}}]{anto12a}
{Antoniadis}, J., {van Kerkwijk}, M.~H., {Koester}, D., {Freire}, P.~C.~C.,
  {Wex}, N., {Tauris}, T.~M., {Kramer}, M., \& {Bassa}, C.~G. 2012, ArXiv
  e-prints

\bibitem[{{Applegate} \& {Shaham}(1994)}]{appl94a}
{Applegate}, J.~H., \& {Shaham}, J. 1994, \apj, 436, 312

\bibitem[{{Archibald} {et~al.}(2009){Archibald}, {Stairs}, {Ransom}, {Kaspi},
  {Kondratiev}, {Lorimer}, {McLaughlin}, {Boyles}, {Hessels}, {Lynch}, {van
  Leeuwen}, {Roberts}, {Jenet}, {Champion}, {Rosen}, {Barlow}, {Dunlap}, \&
  {Remillard}}]{arch09a}
{Archibald}, A.~M., {et~al.} 2009, Science, 324, 1411

\bibitem[{{Arzoumanian} {et~al.}(1994){Arzoumanian}, {Fruchter}, \&
  {Taylor}}]{arzo94a}
{Arzoumanian}, Z., {Fruchter}, A.~S., \& {Taylor}, J.~H. 1994, \apjl, 426, L85+

\bibitem[{{Boyles} {et~al.}(2012){Boyles}, {Lynch}, {Ransom}, {Stairs},
  {Lorimer}, {McLaughlin}, {Hessels}, {Kaspi}, {Kondratiev}, {Archibald},
  {Berndsen}, {Cardoso}, {Cherry}, {Epstein}, {Karako-Argaman}, {McPhee},
  {Pennucci}, {Roberts}, {Stovall}, \& {van Leeuwen}}]{boyl12a}
{Boyles}, J., {et~al.} 2012, ArXiv e-prints

\bibitem[{{Breeveld} {et~al.}(2011){Breeveld}, {Landsman}, {Holland}, {Roming},
  {Kuin}, \& {Page}}]{bree11a}
{Breeveld}, A.~A., {Landsman}, W., {Holland}, S.~T., {Roming}, P., {Kuin},
  N.~P.~M., \& {Page}, M.~J. 2011, in American Institute of Physics Conference
  Series, Vol. 1358, American Institute of Physics Conference Series, ed.
  {J.~E.~McEnery, J.~L.~Racusin, \& N.~Gehrels}, 373--376

\bibitem[{{Breton} {et~al.}(2011){Breton}, {Rappaport}, {van Kerkwijk}, \&
  {Carter}}]{bret11a}
{Breton}, R.~P., {Rappaport}, S.~A., {van Kerkwijk}, M.~H., \& {Carter}, J.~A.
  2011, ArXiv e-prints

\bibitem[{{Callanan} {et~al.}(1995){Callanan}, {van Paradijs}, \&
  {Rengelink}}]{call95a}
{Callanan}, P.~J., {van Paradijs}, J., \& {Rengelink}, R. 1995, \apj, 439, 928

\bibitem[{{Claret}(2001)}]{clar01a}
{Claret}, A. 2001, \mnras, 327, 989

\bibitem[{{Cognard} {et~al.}(2011){Cognard}, {Guillemot}, {Johnson}, {Smith},
  {Venter}, {Harding}, {Wolff}, {Cheung}, {Donato}, {Abdo}, {Ballet}, {Camilo},
  {Desvignes}, {Dumora}, {Ferrara}, {Freire}, {Grove}, {Johnston}, {Keith},
  {Kramer}, {Lyne}, {Michelson}, {Parent}, {Ransom}, {Ray}, {Romani}, {Saz
  Parkinson}, {Stappers}, {Theureau}, {Thompson}, {Weltevrede}, \&
  {Wood}}]{cogn11a}
{Cognard}, I., {et~al.} 2011, \apj, 732, 47

\bibitem[{{Cordes} \& {Lazio}(2002)}]{cord02a}
{Cordes}, J.~M., \& {Lazio}, T.~J.~W. 2002, ArXiv Astrophysics e-prints

\bibitem[{{Dhillon} {et~al.}(2007){Dhillon}, {Marsh}, {Stevenson}, {Atkinson},
  {Kerry}, {Peacocke}, {Vick}, {Beard}, {Ives}, {Lunney}, {McLay}, {Tierney},
  {Kelly}, {Littlefair}, {Nicholson}, {Pashley}, {Harlaftis}, \&
  {O'Brien}}]{dhil07a}
{Dhillon}, V.~S., {et~al.} 2007, \mnras, 378, 825

\bibitem[{{Eichler} \& {Gedalin}(1995)}]{eich95a}
{Eichler}, D., \& {Gedalin}, M. 1995, in Astronomical Society of the Pacific
  Conference Series, Vol.~72, Millisecond Pulsars. A Decade of Surprise, ed.
  {A.~S.~Fruchter, M.~Tavani, \& D.~C.~Backer}, 235

\bibitem[{{Eichler} \& {Levinson}(1988)}]{eich88a}
{Eichler}, D., \& {Levinson}, A. 1988, \apjl, 335, L67

\bibitem[{{Freire} {et~al.}(2011){Freire}, {Bassa}, {Wex}, {Stairs},
  {Champion}, {Ransom}, {Lazarus}, {Kaspi}, {Hessels}, {Kramer}, {Cordes},
  {Verbiest}, {Podsiadlowski}, {Nice}, {Deneva}, {Lorimer}, {Stappers},
  {McLaughlin}, \& {Camilo}}]{frei11a}
{Freire}, P.~C.~C., {et~al.} 2011, \mnras, 412, 2763

\bibitem[{{Fruchter} {et~al.}(1988{\natexlab{a}}){Fruchter}, {Gunn}, {Lauer},
  \& {Dressler}}]{fruc88a}
{Fruchter}, A.~S., {Gunn}, J.~E., {Lauer}, T.~R., \& {Dressler}, A.
  1988{\natexlab{a}}, \nat, 334, 686

\bibitem[{{Fruchter} {et~al.}(1988{\natexlab{b}}){Fruchter}, {Stinebring}, \&
  {Taylor}}]{fruc88b}
{Fruchter}, A.~S., {Stinebring}, D.~R., \& {Taylor}, J.~H. 1988{\natexlab{b}},
  \nat, 333, 237

\bibitem[{{Gaensler} {et~al.}(2008){Gaensler}, {Madsen}, {Chatterjee}, \&
  {Mao}}]{gaen08a}
{Gaensler}, B.~M., {Madsen}, G.~J., {Chatterjee}, S., \& {Mao}, S.~A. 2008,
  Publications of the Astronomical Society of Australia, 25, 184

\bibitem[{{Hessels}(2008)}]{hess08b}
{Hessels}, J.~W.~T. 2008, in American Institute of Physics Conference Series,
  Vol. 1068, American Institute of Physics Conference Series, ed. {R.~Wijnands,
  D.~Altamirano, P.~Soleri, N.~Degenaar, N.~Rea, P.~Casella, A.~Patruno, \&
  M.~Linares}, 130--134

\bibitem[{{Hessels} {et~al.}(2011){Hessels}, {Roberts}, {McLaughlin}, {Ray},
  {Bangale}, {Ransom}, {Kerr}, {Camilo}, \& {Decesar}}]{hess11a}
{Hessels}, J.~W.~T., {et~al.} 2011, in American Institute of Physics Conference
  Series, Vol. 1357, American Institute of Physics Conference Series, ed.
  {M.~Burgay, N.~D'Amico, P.~Esposito, A.~Pellizzoni, \& A.~Possenti }, 40--43

\bibitem[{{Hook} {et~al.}(2004){Hook}, {J{\o}rgensen}, {Allington-Smith},
  {Davies}, {Metcalfe}, {Murowinski}, \& {Crampton}}]{hook04a}
{Hook}, I.~M., {J{\o}rgensen}, I., {Allington-Smith}, J.~R., {Davies}, R.~L.,
  {Metcalfe}, N., {Murowinski}, R.~G., \& {Crampton}, D. 2004, \pasp, 116, 425

\bibitem[{{Jordi} {et~al.}(2006){Jordi}, {Grebel}, \& {Ammon}}]{jord06a}
{Jordi}, K., {Grebel}, E.~K., \& {Ammon}, K. 2006, \aap, 460, 339

\bibitem[{{Kerr} {et~al.}(2012){Kerr}, {Camilo}, {Johnson}, {Ferrara},
  {Guillemot}, {Harding}, {Hessels}, {Johnston}, {Keith}, {Kramer}, {Ransom},
  {Ray}, {Reynolds}, {Sarkissian}, \& {Wood}}]{kerr12a}
{Kerr}, M., {et~al.} 2012, \apjl, 748, L2

\bibitem[{{Kluzniak} {et~al.}(1988){Kluzniak}, {Ruderman}, {Shaham}, \&
  {Tavani}}]{kluz88a}
{Kluzniak}, W., {Ruderman}, M., {Shaham}, J., \& {Tavani}, M. 1988, \nat, 334,
  225

\bibitem[{{Lazaridis} {et~al.}(2011){Lazaridis}, {Verbiest}, {Tauris},
  {Stappers}, {Kramer}, {Wex}, {Jessner}, {Cognard}, {Desvignes}, {Janssen},
  {Purver}, {Theureau}, {Bassa}, \& {Smits}}]{laza11a}
{Lazaridis}, K., {et~al.} 2011, \mnras, 414, 3134

\bibitem[{{Lin} {et~al.}(2011){Lin}, {Rappaport}, {Podsiadlowski}, {Nelson},
  {Paxton}, \& {Todorov}}]{lin11a}
{Lin}, J., {Rappaport}, S., {Podsiadlowski}, P., {Nelson}, L., {Paxton}, B., \&
  {Todorov}, P. 2011, \apj, 732, 70

\bibitem[{{Lorimer} \& {Kramer}(2004)}]{lori04a}
{Lorimer}, D.~R., \& {Kramer}, M. 2004, {Handbook of Pulsar Astronomy}, ed.
  R.~{Ellis}, J.~{Huchra}, S.~{Kahn}, G.~{Rieke}, \& P.~B. {Stetson}

\bibitem[{{Lupton}(2005)}]{lupt05a}
{Lupton}, R.~C. 2005, Transformations between SDSS magnitudes and UBVRcIc

\bibitem[{{Lynch} {et~al.}(2012){Lynch}, {Boyles}, {Ransom}, {Stairs},
  {Lorimer}, {McLaughlin}, {Hessels}, {Kaspi}, {Kondratiev}, {Archibald},
  {Berndsen}, {Cardoso}, {Cherry}, {Karako-Argaman}, {van Leeuwen}, {McPhee},
  {Pennucci}, \& {Roberts}}]{lync12a}
{Lynch}, R.~S., {et~al.} 2012, ArXiv e-prints

\bibitem[{{Manchester} {et~al.}(2005){Manchester}, {Hobbs}, {Teoh}, \&
  {Hobbs}}]{manc05a}
{Manchester}, R.~N., {Hobbs}, G.~B., {Teoh}, A., \& {Hobbs}, M. 2005, \aj, 129,
  1993

\bibitem[{{Mu{\~n}oz-Darias} {et~al.}(2009){Mu{\~n}oz-Darias}, {Casares},
  {O'Brien}, {Steeghs}, {Mart{\'{\i}}nez-Pais}, {Cornelisse}, \&
  {Charles}}]{muno09b}
{Mu{\~n}oz-Darias}, T., {Casares}, J., {O'Brien}, K., {Steeghs}, D.,
  {Mart{\'{\i}}nez-Pais}, I.~G., {Cornelisse}, R., \& {Charles}, P.~A. 2009,
  \mnras, 394, L136

\bibitem[{{Poole} {et~al.}(2008){Poole}, {Breeveld}, {Page}, {Landsman},
  {Holland}, {Roming}, {Kuin}, {Brown}, {Gronwall}, {Hunsberger}, {Koch},
  {Mason}, {Schady}, {vanden Berk}, {Blustin}, {Boyd}, {Broos}, {Carter},
  {Chester}, {Cucchiara}, {Hancock}, {Huckle}, {Immler}, {Ivanushkina},
  {Kennedy}, {Marshall}, {Morgan}, {Pandey}, {de Pasquale}, {Smith}, \&
  {Still}}]{pool08a}
{Poole}, T.~S., {et~al.} 2008, \mnras, 383, 627

\bibitem[{{Portegies Zwart} {et~al.}(2011){Portegies Zwart}, {van den Heuvel},
  {van Leeuwen}, \& {Nelemans}}]{port11a}
{Portegies Zwart}, S., {van den Heuvel}, E.~P.~J., {van Leeuwen}, J., \&
  {Nelemans}, G. 2011, \apj, 734, 55

\bibitem[{{Ransom} {et~al.}(2011){Ransom}, {Ray}, {Camilo}, {Roberts}, {{\c
  C}elik}, {Wolff}, {Cheung}, {Kerr}, {Pennucci}, {DeCesar}, {Cognard}, {Lyne},
  {Stappers}, {Freire}, {Grove}, {Abdo}, {Desvignes}, {Donato}, {Ferrara},
  {Gehrels}, {Guillemot}, {Gwon}, {Harding}, {Johnston}, {Keith}, {Kramer},
  {Michelson}, {Parent}, {Saz Parkinson}, {Romani}, {Smith}, {Theureau},
  {Thompson}, {Weltevrede}, {Wood}, \& {Ziegler}}]{rans11a}
{Ransom}, S.~M., {et~al.} 2011, \apjl, 727, L16

\bibitem[{{Ray} {et~al.}(2012){Ray}, {Abdo}, {Parent}, {Bhattacharya},
  {Bhattacharyya}, {Camilo}, {Cognard}, {Theureau}, {Ferrara}, {Harding},
  {Thompson}, {Freire}, {Guillemot}, {Gupta}, {Roy}, {Hessels}, {Johnston},
  {Keith}, {Shannon}, {Kerr}, {Michelson}, {Romani}, {Kramer}, {McLaughlin},
  {Ransom}, {Roberts}, {Saz Parkinson}, {Ziegler}, {Smith}, {Stappers},
  {Weltevrede}, \& {Wood}}]{ray12b}
{Ray}, P.~S., {et~al.} 2012, ArXiv e-prints

\bibitem[{{Reynolds} {et~al.}(2007){Reynolds}, {Callanan}, {Fruchter},
  {Torres}, {Beer}, \& {Gibbons}}]{reyn07a}
{Reynolds}, M.~T., {Callanan}, P.~J., {Fruchter}, A.~S., {Torres}, M.~A.~P.,
  {Beer}, M.~E., \& {Gibbons}, R.~A. 2007, \mnras, 379, 1117

\bibitem[{{Roberts}(2011)}]{robe11a}
{Roberts}, M.~S.~E. 2011, in American Institute of Physics Conference Series,
  Vol. 1357, American Institute of Physics Conference Series, ed. M.~{Burgay},
  N.~{D'Amico}, P.~{Esposito}, A.~{Pellizzoni}, \& A.~{Possenti}, 127--130

\bibitem[{{Roberts}(2012)}]{robe12a}
{Roberts}, M.~S.~E. 2012, ArXiv e-prints

\bibitem[{{Romani} {et~al.}(2012){Romani}, {Filippenko}, {Silverman}, {Cenko},
  {Greiner}, {Rau}, {Elliott}, \& {Pletsch}}]{roma12b}
{Romani}, R.~W., {Filippenko}, A.~V., {Silverman}, J.~M., {Cenko}, S.~B.,
  {Greiner}, J., {Rau}, A., {Elliott}, J., \& {Pletsch}, H.~J. 2012, \apjl,
  760, L36

\bibitem[{{Romani} \& {Shaw}(2011)}]{roma11a}
{Romani}, R.~W., \& {Shaw}, M.~S. 2011, \apjl, 743, L26

\bibitem[{{Schlegel} {et~al.}(1998){Schlegel}, {Finkbeiner}, \&
  {Davis}}]{schl98a}
{Schlegel}, D.~J., {Finkbeiner}, D.~P., \& {Davis}, M. 1998, \apj, 500, 525

\bibitem[{{Stappers} {et~al.}(2003){Stappers}, {Gaensler}, {Kaspi}, {van der
  Klis}, \& {Lewin}}]{stap03a}
{Stappers}, B.~W., {Gaensler}, B.~M., {Kaspi}, V.~M., {van der Klis}, M., \&
  {Lewin}, W.~H.~G. 2003, Science, 299, 1372

\bibitem[{{Stappers} {et~al.}(2001){Stappers}, {van Kerkwijk}, {Bell}, \&
  {Kulkarni}}]{stap01b}
{Stappers}, B.~W., {van Kerkwijk}, M.~H., {Bell}, J.~F., \& {Kulkarni}, S.~R.
  2001, \apjl, 548, L183

\bibitem[{{Stappers} {et~al.}(1996){Stappers}, {Bailes}, {Lyne}, {Manchester},
  {D'Amico}, {Tauris}, {Lorimer}, {Johnston}, \& {Sandhu}}]{stap96b}
{Stappers}, B.~W., {et~al.} 1996, \apjl, 465, L119+

\bibitem[{{Tavani}(1993)}]{tava93a}
{Tavani}, M. 1993, \aaps, 97, 313

\bibitem[{{van den Heuvel} \& {van Paradijs}(1988)}]{van-88a}
{van den Heuvel}, E.~P.~J., \& {van Paradijs}, J. 1988, \nat, 334, 227

\bibitem[{{van Kerkwijk} {et~al.}(2011){van Kerkwijk}, {Breton}, \&
  {Kulkarni}}]{van-11a}
{van Kerkwijk}, M.~H., {Breton}, R.~P., \& {Kulkarni}, S.~R. 2011, \apj, 728,
  95

\bibitem[{{van Kerkwijk} \& {Ingle}(2008)}]{van-08a}
{van Kerkwijk}, M.~H., \& {Ingle}, A. 2008, \apjl, 683, L159

\bibitem[{{van Paradijs} {et~al.}(1988){van Paradijs}, {Allington-Smith},
  {Callanan}, {Charles}, {Hassall}, {Machin}, {Mason}, {Naylor}, \&
  {Smale}}]{van-88b}
{van Paradijs}, J., {et~al.} 1988, \nat, 334, 684

\bibitem[{{Wang} {et~al.}(2011){Wang}, {Breton}, {Heinke}, {Deloye}, \&
  {Zhong}}]{wang11a}
{Wang}, Z., {Breton}, R.~P., {Heinke}, C.~O., {Deloye}, C.~J., \& {Zhong}, J.
  2011, ArXiv e-prints

\end{thebibliography}

\clearpage

\begin{table}[!ht]
\begin{center}
\caption{Photometry of PSR~J0023+09}
\begin{tabular}{cccccc}
\hline
\hline
Time & Orbital Phase\tablenotemark{a} & Flux & Flux Error\tablenotemark{c} & Magnitude\tablenotemark{b} & Magnitude Error\tablenotemark{c} \\
(MJD) &  & ($\mu Jy$) & ($\mu Jy$) &  & \\
\hline
\multicolumn{6}{c}{$g$-band, GMOS} \\
\hline
55445.513890 & 0.6410 & 0.989 & 0.065 & 23.908 & 0.070 \\
55449.575790 & 0.9056 & 0.024 & 0.057 & 26.756 & 0.511 \\
55449.587001 & 0.9863 & 0.001 & 0.065 & 26.964 & 0.599 \\
55449.612403 & 0.1693 & 0.106 & 0.062 & 26.099 & 0.428 \\
\hline
\multicolumn{6}{c}{$i$-band, GMOS} \\
\hline
55445.507793 & 0.5970 & 6.397 & 0.252 & 21.883 & 0.043 \\
55445.519981 & 0.6848 & 4.368 & 0.224 & 22.296 & 0.055 \\
55449.569694 & 0.8616 & 0.956 & 0.239 & 23.889 & 0.244 \\
55449.581881 & 0.9494 & 0.637 & 0.240 & 24.266 & 0.330 \\
55449.606307 & 0.1254 & 1.715 & 0.238 & 23.295 & 0.146 \\
55449.618493 & 0.2132 & 3.027 & 0.242 & 22.691 & 0.086 \\
\end{tabular}\label{t:j0023}
\tablenotetext{1}{Orbital phases are measured from the companion's inferior conjunction.}
\tablenotetext{2}{AB magnitudes in the Lupton system, $m = m_{_0} - 2.5 \log b^\prime - (2.5 \log e) \sinh^{-1} (f/2b^\prime)$, using softening parameters $b^\prime = 0.059$ and 0.233 $\mu Jy$ for the $g$ and $i$ band, respectively, and zero-point $m_{_0} = -48.6$.}
\tablenotetext{3}{The flux and magnitude errors represent the formal uncertainties. One should also add in quadrature the zero-point calibration errors, which are 0.319 and 0.757 mag in $g$ and $i$ band, respectively.}
\end{center}
\end{table}

\begin{table}[!ht]
\begin{center}
\caption{Photometry of PSR~J2256$-$10}
\begin{tabular}{cccccc}
\hline
\hline
Time & Orbital Phase\tablenotemark{a} & Flux & Flux Error\tablenotemark{c} & Magnitude\tablenotemark{b} & Magnitude Error\tablenotemark{c} \\
(MJD) &  & ($\mu Jy$) & ($\mu Jy$) &  & \\
\hline
\multicolumn{6}{c}{$g$-band, GMOS} \\
\hline
55428.546679 & 0.2850 & 1.871 & 0.060 & 23.219 & 0.035 \\
55449.283273 & 0.6936 & 1.835 & 0.029 & 23.241 & 0.017 \\
55454.418742 & 0.8170 & 0.284 & 0.026 & 25.256 & 0.098 \\
55469.350249 & 0.9567 & 0.020 & 0.024 & 27.478 & 0.483 \\
\hline
\multicolumn{6}{c}{$i$-band, GMOS} \\
\hline
55428.540594 & 0.2564 & 6.770 & 0.130 & 21.823 & 0.021 \\
55428.552779 & 0.3136 & 10.368 & 0.125 & 21.360 & 0.013 \\
55449.277175 & 0.6649 & 9.902 & 0.088 & 21.410 & 0.010 \\
55449.291633 & 0.7328 & 5.578 & 0.069 & 22.033 & 0.013 \\
55454.412643 & 0.7884 & 3.321 & 0.078 & 22.596 & 0.026 \\
55454.424831 & 0.8456 & 1.559 & 0.074 & 23.416 & 0.051 \\
55469.344150 & 0.9280 & 0.867 & 0.067 & 24.049 & 0.083 \\
55469.356340 & 0.9853 & 0.816 & 0.062 & 24.114 & 0.081 \\
\hline
\multicolumn{6}{c}{$g$-band, ULTRACAM} \\
\hline
55323.421265 & 0.4664 & 3.979 & 1.606 & 22.245 & 0.335 \\
\hline
\multicolumn{6}{c}{$z$-band, ULTRACAM} \\
\hline
55323.421265 & 0.4664 & 45.700 & 5.609 & 19.733 & 0.129 \\
\end{tabular}\label{t:j2256}
\tablenotetext{1}{Orbital phases are measured from the companion's inferior conjunction.}
\tablenotetext{2}{AB magnitudes in the Lupton system, $m = m_{_0} - 2.5 \log b^\prime - (2.5 \log e) \sinh^{-1} (f/2b^\prime)$, using softening parameters $b^\prime = 0.025$, 0.064, 1.673 and 5.845 $\mu Jy$ for the $g$ (GMOS), $i$, $g$ (ULTRACAM) and $z$ band, respectively, and zero-point $m_{_0} = -48.6$.}
\tablenotetext{3}{The flux and magnitude errors represent the formal uncertainties. One should also add in quadrature the zero-point calibration errors, which are 0.104, 0.119, 0.010 and 0.010 mag in $g$ (GMOS), $i$, $g$ (ULTRACAM) and $z$ band, respectively.}
\end{center}
\end{table}

\begin{table}[!ht]
\begin{center}
\caption{Photometry of PSR~J1810+17}
\begin{tabular}{cccccc}
\hline
\hline
Time & Orbital Phase\tablenotemark{a} & Flux & Flux Error\tablenotemark{c} & Magnitude\tablenotemark{b} & Magnitude Error\tablenotemark{c} \\
(MJD) &  & ($\mu Jy$) & ($\mu Jy$) &  & \\
\hline
\multicolumn{6}{c}{$g$-band, GMOS} \\
\hline
55442.236873 & 0.7092 & 40.430 & 0.119 & 19.883 & 0.003 \\
55443.292770 & 0.8355 & 11.639 & 0.053 & 21.235 & 0.005 \\
55446.293441 & 0.0870 & 6.537 & 0.061 & 21.861 & 0.010 \\
55447.310477 & 0.9509 & 1.066 & 0.049 & 23.828 & 0.050 \\
\hline
\multicolumn{6}{c}{$i$-band, GMOS} \\
\hline
55442.230765 & 0.6680 & 42.101 & 0.141 & 19.839 & 0.004 \\
55442.242953 & 0.7502 & 29.091 & 0.124 & 20.240 & 0.005 \\
55443.286671 & 0.7943 & 20.585 & 0.132 & 20.616 & 0.007 \\
55443.298861 & 0.8766 & 8.369 & 0.107 & 21.593 & 0.014 \\
55446.287344 & 0.0458 & 5.916 & 0.120 & 21.969 & 0.022 \\
55446.299530 & 0.1281 & 14.863 & 0.126 & 20.969 & 0.009 \\
55447.304378 & 0.9098 & 5.780 & 0.095 & 21.995 & 0.018 \\
55447.316567 & 0.9920 & 3.715 & 0.107 & 22.474 & 0.031 \\
\end{tabular}\label{t:j1810}
\tablenotetext{1}{Orbital phases are measured from the companion's inferior conjunction.}
\tablenotetext{2}{AB magnitudes in the Lupton system, $m = m_{_0} - 2.5 \log b^\prime - (2.5 \log e) \sinh^{-1} (f/2b^\prime)$, using softening parameters $b^\prime = 0.051$ and 0.099 $\mu Jy$ for the $g$ and $i$ band, respectively, and zero-point $m_{_0} = -48.6$.}
\tablenotetext{3}{The flux and magnitude errors represent the formal uncertainties. One should also add in quadrature the zero-point calibration errors, which are 0.301 and 0.308 mag in $g$ and $i$ band, respectively.}
\end{center}
\end{table}

\begin{table}[!ht]
\begin{center}
\caption{Photometry of PSR~J2215+51}
\begin{tabular}{cccccc}
\hline
\hline
Time & Orbital Phase\tablenotemark{a} & Flux & Flux Error\tablenotemark{c} & Magnitude\tablenotemark{b} & Magnitude Error\tablenotemark{c} \\
(MJD) &  & ($\mu Jy$) & ($\mu Jy$) &  & \\
\hline
\multicolumn{6}{c}{$g$-band, GMOS} \\
\hline
55531.246057 & 0.1970 & 82.000 & 0.232 & 19.115 & 0.003 \\
\hline
\multicolumn{6}{c}{$i$-band, GMOS} \\
\hline
55531.192316 & 0.8855 & 59.559 & 1.180 & 19.462 & 0.022 \\
55531.238446 & 0.1529 & 71.956 & 0.607 & 19.257 & 0.009 \\
55531.252146 & 0.2323 & 95.277 & 0.448 & 18.952 & 0.005 \\
55531.256359 & 0.2567 & 102.023 & 0.443 & 18.878 & 0.005 \\
\hline
\multicolumn{6}{c}{uvw1-band, UVOT} \\
\hline
55399.269193 & 0.1227 & 0.670 & 2.759 & 22.979 & 0.749 \\
55399.335860 & 0.5092 & 4.868 & 2.407 & 22.036 & 0.417 \\
55399.402162 & 0.8936 & -1.056 & 2.269 & 23.450 & 0.603 \\
55399.536221 & 0.6707 & 2.450 & 2.136 & 22.525 & 0.499 \\
55399.602885 & 0.0572 & -1.364 & 2.157 & 23.531 & 0.561 \\
55399.665753 & 0.4216 & 5.503 & 2.026 & 21.930 & 0.325 \\
55399.730436 & 0.7966 & 2.269 & 2.144 & 22.568 & 0.511 \\
55399.799357 & 0.1961 & -0.123 & 1.893 & 23.196 & 0.521 \\
55399.865723 & 0.5808 & 5.094 & 2.032 & 21.997 & 0.342 \\
55399.933775 & 0.9753 & -0.077 & 1.976 & 23.183 & 0.544 \\
\end{tabular}\label{t:j2215}
\tablenotetext{1}{Orbital phases are measured from the companion's inferior conjunction.}
\tablenotetext{2}{AB magnitudes in the Lupton system, $m = m_{_0} - 2.5 \log b^\prime - (2.5 \log e) \sinh^{-1} (f/2b^\prime)$, using softening parameters $b^\prime = 0.242$, 0.462 and 1.972 $\mu Jy$ for the $g$, $i$ and uvw1 band, respectively, and zero-point $m_{_0} = -48.6$.}
\tablenotetext{3}{The flux and magnitude errors represent the formal uncertainties. One should also add in quadrature the zero-point calibration errors, which are 0.302 and 0.302 mag in $g$ and $i$ band, respectively.}
\end{center}
\end{table}

\end{document}